\documentclass[prd,twocolumn]{revtex4}
\usepackage{graphicx}

\newcommand\lapp{\mathrel{\rlap{\lower4pt\hbox{\hskip1pt$\sim$}}
    \raise1pt\hbox{$<$}}}
\newcommand\gapp{\mathrel{\rlap{\lower4pt\hbox{\hskip1pt$\sim$}}
    \raise1pt\hbox{$>$}}}
\newcommand\eapp{\mathrel{\rlap{\raise2pt\hbox{\hskip0pt$\sim$}}
    \lower1pt\hbox{$-$}}}

\begin{document}
\newcommand{\be}{\begin{equation}}
\newcommand{\ee}{\end{equation}}
\newcommand{\ben}{\begin{eqnarray}}
\newcommand{\een}{\end{eqnarray}}
\newcommand{\bb}{\bibitem}
\newcommand{\ov}{\overline}
\newcommand{\wt}{\widetilde}
\newcommand{\nn}{\nonumber}

\title[Varying Alpha Monopoles]{Varying Alpha Monopoles}

\author{J. Menezes\footnote{jmenezes@fc.up.pt}, P. P. Avelino\footnote{ppavelin@fc.up.pt}, and C. Santos\footnote{cssilva@fc.up.pt}}

\address{Centro de F\'{\i}sica do Porto e Departamento de F\'{\i}sica
da Faculdade de Ci\^encias da Universidade do Porto, Rua do Campo Alegre 687,
4169-007, Porto, Portugal}

\begin{abstract}

We study static magnetic monopoles in the context of varying-$\alpha$ 
theories and show that there is a group of models for which the 
t'Hooft-Polyakov solution is still valid. Nevertheless, in general 
static magnetic monopole solutions in varying-$\alpha$ theories depart 
from the classical t'Hooft-Polyakov solution with the electromagnetic 
energy concentrated inside the core seeding spatial variations of 
the fine structure constant.
We show that Equivalence Principle 
constraints impose tight limits on the allowed
variations of $\alpha$ induced by magnetic monopoles which confirms 
the difficulty to generate significant large-scale spatial variation of 
the fine structure constant found in previous works. This is true even in the 
most favorable case where magnetic monopoles are the source for these variations.

\end{abstract}
\maketitle

\section{Introduction}

The possibility that at least some of the fundamental ``constants'' of 
Nature might be dynamical appears naturally in the context of models 
with extra-spatial dimensions. The interest in this type of models 
has recently been increased with results coming from both quasar 
absorption systems \cite{webb,murphy} 
(see however \cite{quast,chand,srianand}) and 
the Oklo natural nuclear reactor \cite{fujii} suggesting a cosmological
variation of the fine-structure constant, 
$\alpha=e^2/(4\,\pi\,\hbar\, c)$ at low red-shifts. Other constraints at 
low redshift include atomic clocks \cite{marion} and meteorites 
\cite{olive} while at high
redshifts there are also upper limits to the allowed variations of $\alpha$
coming from either the Cosmic Microwave Background or
Big Bang Nucleossynthesis 
\cite{hannestad,avelino,martins,rocha,martins2,sigurdson,rocha2}.

On the theoretical side some effort has been made on the 
construction of self-consistent phenomenological models for space-time 
$\alpha$ variability most of them based on the original Bekenstein model 
\cite{bekenstein}. In some of these models 
\cite{avelino2,lee,anchordoqui,parkinson,copeland,nunes,sandvikx,doranx} the 
possibility that the variation of fine structure constant might be 
coupled to the dark energy equation of state responsible for the recent 
acceleration of the Universe \cite{tonry,bennett} has also been explored.

Most of these studies have focused mainly on the 
variation of $\alpha$ with time. The reason for this is that 
it is very difficult to generate significant large-scale spatial variations 
of the fine structure constant. This has been shown in \cite{menezes} 
(see also \cite{magueijo}) in a model where the spatial variation of 
$\alpha$ were induced by cosmic strings cosmic and in \cite{barrow} in a more 
general context. In the cosmic string case, the electromagnetic energy 
concentrated along the core of the local string is a source of these 
spatial variations, which are roughly proportional to the gravitational 
potential induced by the strings. They are constrained by Equivalence Principle to be small and 
overall limits for the allowed spatial variations have been calculated. 

In this article, we consider the case of spatial variations of $\alpha$ 
seeded by a static non-Abelian local monopole. We show that the 
electromagnetic energy in its core can produce spatial variations of the 
fine-structure constant in the vicinity of the monopole in the context of 
Bekenstein-type models. 
We start by solving the standard problem with no $\alpha$ variation and 
recover the classical solution presented by t'Hooft and Polyakov 
\cite{thooft,polyakov} (see for instance Refs. \cite{kirkman,vilenkin,manton,forgacs}) 
and then proceed to investigate other solutions with variable $\alpha$.

The article is organized as 
follows. In Sec. II we introduce Bekenstein-type models in Yang-Mills theories 
and obtain the equations that describe a static magnetic monopole. We write the 
energy density of the monopole and show that the standard 
electromagnetism is recovered outside the core if $\alpha$ is a constant. In. Sec. III 
we present the 
numerical technique applied to solve the equations of motion and give 
results for several choices of the gauge kinetic 
function. In Sec. IV we study the limits imposed by the Equivalence Principle 
on the allowed variations of $\alpha$ in the vicinity of the monopole, as a 
function of the symmetry breaking scale. Finally, in Sec. IV, we summarize 
and discuss our results. Throughout this paper we shall use units in which 
$\hbar = c = 1$ and the metric signature $+---$.


\section{Bekenstein-type Models in Non-Abelian Field Theories}


We first introduce Bekenstein-type models in a non-Abelian Yang-Mills theory. We take the electric charge to be a function of the space-time coordinates, $e=e_0\,\epsilon(x^\mu)$ in which $\epsilon$ is a real scalar field and $e_0$ is an arbitrary constant charge. 
Let the Higgs field $\Phi^{a}$ be an isovector, where $a=1,2,3$ are internal indices associated to isospace
with $SU(2)$ symmetry.

The Lagrangean density is given by
\ben
\mathcal{L}\,&=&\, \frac12\,\left(D_\mu\,\Phi^a \right)\left(D^\mu\,\Phi^a\right) - V(\Phi^a) \nn \\
&-&\frac{B_F(\varphi)}{4}\,f_{\mu\nu}^{a}\,f^{a\mu\nu}
+\frac12\,\partial_\mu\,\varphi\,\partial^\mu\,\varphi, \label{laga}
\een 
where 
\be
B_F(\varphi)=\epsilon(\varphi)^{-2} \label{defini}
\ee
is the gauge kinetic function of a massless scalar field $\varphi$. 
This function acts as the effective dielectric 
permittivity and can phenomenologically be taken as an arbitrary function of $\varphi$.
 
Defining an auxiliary gauge field $a_\mu^a=\epsilon\,A_\mu^a$ and a new non-Abelian gauge field strength by
\be
f_{\mu\nu}^a=\,\epsilon\,F_{\mu\nu}^a=\partial_\mu a_\nu^a - \partial_\nu a_\mu^a + e_0 \epsilon^{a\,b\,c}\,a_{\mu}^b a_{\nu}^c, \label{gfs}
\ee
the covariant derivatives are written in the usual form 
\ben
D_\mu \Phi^a &=& \partial_\mu \Phi^a + e_0\,\epsilon^{a\,b\,c} a_\mu^b\,\Phi^c, \label{cd}
\een
where $\epsilon^{a\,b\,c}$ is the Levi-Civita tensor.


We consider that the potential $V(\Phi^a)$ is given by
\be
V(\Phi^a) = \frac{\gamma}{8}\,\left(\Phi^a\,\Phi^a-v^2\right)^2
\ee
where $\gamma > 0$ is the coupling of the scalar self-interaction, and $v$ is the vacuum
expectation value of the Higgs field. 
The Lagrangian density in eqn. (\ref{laga}) is then invariant 
under SU(2) gauge transformations of the form 
\ben
\delta\, \Phi^a &=& \epsilon^{a\,b\,c} \Phi^b \Lambda^c,\\
\delta\, a_\mu^a&=& \epsilon^{a\,b\,c} a_{\mu}^b \, \Lambda^c + e_0 \partial_\mu\,\Lambda^a,
\een
where $\Lambda^a$ is a generic isovector. 
This symmetry is broken down to $U(1)$ because there is a non vanishing expectation value of the Higgs field. Thus the two components of the vector field develop a mass $M_W = v\,e_0$, while
the mass of the Higgs field $M_H = v\,\sqrt{\gamma}$.

It is convenient to define the dimensionless ratio
\be
\zeta\,=\,\frac{M_H}{M_W}\,=\frac{\sqrt{\gamma}}{e_0},
\ee
and to rescale the radial coordinate by $M_W$, so that distance is expressed in units of $M_W^{-1}$.

Varying the action with respect to the adjoint Higgs field $\Phi^{a\dag}$ one gets
\be\label{eqphi}
\left( D_\mu\,D^\mu\,\Phi^a \right)\,=\,-\gamma\,\Phi^a\,\left(\Phi^b\,\Phi^b - v^2 \right).
\ee
Variation with respect to $a_\mu^a$ leads to
\be\label{eqfmn}
D_\nu\,\left[ B_F(\varphi) f^{a\,\mu\nu} \right] = j^{a\nu}
\ee
with the current $j^{a\mu}$ defined as
\be
j^{a\mu} =  e_0\,\epsilon^{a\,b\,c} \,\Phi^b\, (D^\mu\,\Phi^c).
\ee
Finally, variation with respect to $\varphi$ gives
\be\label{eqbox}
\partial_\mu\,\partial^\mu\,\varphi\,=\,-\frac{1}{4}\,
\frac{\partial B_F(\varphi)}{\partial\, \varphi}\, f^2,\\
\ee
in which $f^2\,=\,f^a_{\mu\nu}\,f^{a\,\mu\nu}$.

We are now interested in static, spherically symmetric, magnetic monopole solutions. Therefore we make the $``$hedgehog$"$ ansatz 
\ben
\Phi^a(r)&=&H(r)\,\frac{x^a}{r},\label{Hig}\\
a_0^a(r)&=&0,\label{a0}\\
a_i^a(r)&=&\epsilon_{iak}\,\frac{x_k}{e_0\,r^2}\,[W(r)-1],\label{ai}
\een
where $x^a$ are the Cartesian coordinates and $r^2=x^k\,x_k$. $H(r)$ and $W(r)$ are dimensionless radial functions which
minimize the self-energy, i.e., the mass of the monopole
\ben
E&=&\frac{4\pi v}{e_0}\int_{0}^{\infty}dr\{\frac{r^2}{2}\left(\frac{dH}{dr}\right)^2 + H^2 W^2 + \frac{\zeta^2 r^2}{8}(1-H^2)^2 \nn\\
&+&B_F(\varphi)\left[\left(\frac{dW}{dr}\right)^2 + \frac{(1-W^2)^2}{2r^2}\right] + \frac{r^2}{2}\left(\frac{d\varphi}{dr}\right)^2 \}, \label{act}
\een
where the coordinate $r$ and the functions $H$ and $\varphi$ have been rescaled as
\be
r\,\rightarrow\,\frac{r}{e_0\,v},\,\,\,\,\,\,\,\,\,\,H\,\rightarrow\,v\,H,\,\,\,\,\,\,\,\,\,\,\varphi\,\rightarrow\,v\,\varphi.
\ee
It will prove useful to compute the energy density which is given by
\ben
\rho &=& \frac{v}{e_0} B_F(\varphi)\left[\left(\frac{W'}{r} \right)^2 + \frac12 \left(\frac{1-W^2}{r^2}\right)^2 \right]\nn \\
&+& \frac{v}{e_0} \left[ \frac{H'^{2}}{2}  + \left(\frac{WH}{r} \right)^2  + \frac{\zeta^2}{8}(1-H^2)^2 + \frac12\,\varphi'^{2}\right] \label{ro}
\een
with a prime meaning a derivative with respect to the dimensionless coordinate $r$.
Substituting the ansatz given in (\ref{Hig})-(\ref{ai}) into equations (\ref{eqphi})-(\ref{eqbox}) one gets
\ben
&&\frac1{r^2}\,\left(r^2 H' \right)'- \left[\frac{\zeta^2}{2}(1-H^2)+\frac{2W^2}{r^2}\right]\,H=0 \label{one}\\
&&\left[B_F(\varphi)\,W'\,\right]'-\left[\frac{B_F(\varphi)}{r^2}(1-W^2) + H^2 \right]W=0 \label{two}\\
&&\frac1{r^2}\left(r^2 \varphi' \right)'-\frac{dB_F(\varphi)}{d\varphi}\left[W'^2+ \frac12\left(\frac{1-W^2}{r^2}\right)^2\right]=0.\label{three}
\een

Defining
\be
f_{\mu\,\nu}=\frac{\Phi^a}{|\Phi|}\,f_{\mu\,\nu}^{a}+\frac1{e_0\,|\Phi|^3}\,\epsilon^{a\,b\,c}\,\Phi^a\,(D_\mu\Phi^b)(D_\nu\Phi^c) \label{fmn}
\ee
and choosing the gauge where $\Phi^a=\delta^{a\,3}\,|\Phi|$, i.e., with $\Phi$ pointing in the same direction everywhere, one gets
\be
f_{\mu\nu}=\partial_\mu\,a_\nu^3-\partial_\nu\,a_\mu^3. \label{f3}
\ee
By writing $a_\mu^3 = a_\mu$, one identifies (\ref{f3}) with the usual electromagnetic tensor which for 
constant $\varphi$ satisfies the ordinary Maxwell equations everywhere except 
in the region where $H \sim 0$.

\section{Numerical Implementation of Equations of Motion}

In order to solve numerically the equations of motion, we have to reduce them to a set of first order equations. For that purpose we define the variables
\be
V\,=\,H'\,\,,\,\,\,\,\,\,\,\,\,\,\,\,U\,=\,W'\,\,,\,\,\,\,\,\,\,\,\,\,\,\,\,\eta\,=\,\varphi'.\label{varr}
\ee			
Equations (\ref{one}-\ref{three}) can then be written as
\ben
\label{IV}
V'&=&-\frac{2V}{r}+\frac{\zeta^2}{2}\,H(H^2-1)+\frac{2\,W^2\,H}{r^2}\\
\label{V}
U'&=&\frac{1}{B_F}\left[-\frac{dB_F}{d\varphi}\eta\,U
+\frac{B_F}{r^2}(W^2-1)W+ H^2W \right]\\
\label{VI}
\eta'&=&\,-\,\frac{2\,\eta}{r}+\frac{dB_F}{d\varphi}\left[U^2+\left(\frac{1-W^2}{r^2}\right)^2\right]
\een
which require at least six boundary conditions.

At far distances from the core ($r \rightarrow \infty$), the Higgs field $H(r)$ falls off to its vacuum value, $H=1$. Using this boundary condition in eqn. (\ref{one}), one gets that $W(r)$ must vanish far from the core. On the other hand, since the symmetry at the core is not broken, then the Higgs field vanishes and from regularity of the energy-momentum tensor, one can choose a gauge for which $W=1$.
The other two boundary conditions come from the normalization of the electric charge at the origin. At the core $e=e_0$, which means that the gauge kinetic function is equal to one. Finally, by using the Gauss law to solve the eqn. (\ref{three}) without sources of $\alpha$ variation other than the monopole, one gets that $\eta$ must vanish at $r=0$. 
Therefore, we have at all six boundary conditions, four of them at origin and the other two far away from the core:
\be
\lim_{r \rightarrow 0} H(r) = 0\,\,\,\,\,\,\,\,\,\,
\lim_{r \rightarrow \infty} H(r) = 1\, ,  \label{condihiggs}
\ee
\be
\lim_{r \rightarrow 0} W(r) = 1\,\,\,\,\,\,\,\,\,\,\,
\lim_{r \rightarrow \infty} W(r) = 0,  \label{condigauge}
\ee
\be
\lim_{r \rightarrow 0} B_F (r) = 1\,\,\,\,\,\,\,\,\,\,\,
\lim_{r \rightarrow 0} \eta(r) =  0\, .  \label{condivarphi}
\ee

As the boundary conditions are at different points of the domain of the functions to be found, we use the relaxation numerical method replacing the set of differential equations by finite-difference equations on a grid of points that covers the whole range of the integration. 

\subsection{t'Hooft-Polyakov Standard Solution}

In this section we consider as a first example the special case of $B_F=1$. This recovers the standard t'Hooft-Polyakov monopole problem \cite{thooft} in which there is no variation of the fine-structure constant with $\varphi(r) = {\rm{constant}}$.

Let us first consider $\zeta=0$, i.e., the case of a massless Higgs field. The energy of the monopole can be written as
\ben
E &=& \frac{4\,\pi}{e_0}\,\int_0^{\infty}\,dr\, \frac12 \left[r\,H'-\frac{(1-W^2)}{r}\right]^2 + \left[W'+W\,H \right]^2\nn\\
&+& \frac{4\,\pi}{e_0}\,\int_0^{\infty}\,d\,P, 
\een 
where we have introduced a new scalar field $P(r)$ as
\be
P(r)=H\,(1-W^2).
\ee
Note that for $H(r)$ and $W(r)$ that solve the first-order differential equations
\ben
H' &=& \frac{1-W^2}{r^2} \label{monof}\\
W' &=& -W\,H \label{monog}
\een
the energy is minimized to
\be
E = 4\,\pi\,\left[P(r \rightarrow \infty) - P(r=0) \right].
\ee

\begin{figure}
\begin{center}
\includegraphics*[width=6cm]{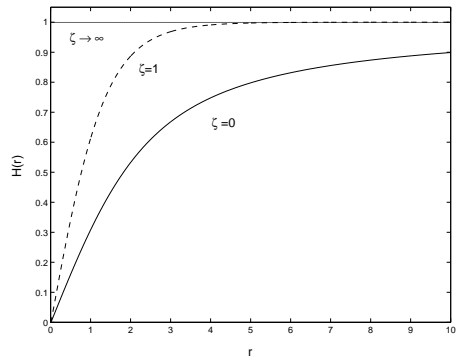}
\end{center}
\caption{
The numerical solution of the field $H(r)$  as a function of 
distance, $r$, to the core of the t'Hooft-Polyakov monopole for several values of $\zeta$. For $\zeta=0$ the Higgs field is massless and for $\zeta \rightarrow \infty$ the Higgs field is frozen at its vacuum value except at the origin.}
\label{kovH}
\end{figure}

\begin{figure}
\begin{center}
\includegraphics*[width=6cm]{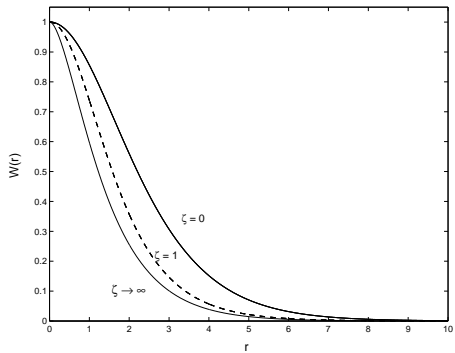}
\end{center}
\caption{
The numerical solution of the field $W(r)$  as a function of 
distance, $r$, to the core of the t'Hooft-Polyakov monopole for three values of $\zeta$.}
\label{kovW}
\end{figure}

This procedure was first presented by Bogomoln'yi \cite{bogomolnyi} and 
generalized in \cite{menezes0,menezes1,menezes2}.
Taking into account the boundary conditions for $H(r)$ and $W(r)$, one gets that the minimum of energy is equal to $4\,\pi/e_0$.
The solutions of equations (\ref{monof},\ref{monog}) are
\ben
H(r) &=& \frac{r}{\sinh{(r)}} \label{bps1}\\
W(r) &=& \frac1{\tanh{(r)}} - \frac1{r} \label{bps2}
\een
which is the well known BPS (Bogomol'nyi, Prasad, Sommerfield) solution \cite{bogomolnyi,prasad}. In Figs. \ref{kovH} and \ref{kovW} we plot $H(r)$ and $W(r)$ given by eqns. (\ref{bps1}) and (\ref{bps2}).

Let us now consider the opposite limit, $\zeta \rightarrow \infty$. Fixing $e_0$ one gets in this limit that $\gamma \rightarrow \infty$, i.e., the Higgs potential is much larger than the kinetic term forcing the Higgs field to be frozen at its vacuum value everywhere except at the origin. Then the only equation of motion is
\be
W''\,=W\,-\frac{W(1-W^2)}{r^2}.
\ee
In Figs. \ref{kovH} and \ref{kovW}, we plot $H(r)$ and $W(r)$ for $\zeta \rightarrow \infty$.

Of course, the former values of $\zeta$ simplify very much the equations of motion. However, we have found numerically the set of solutions for a generic non vanishing finite $\zeta$. In fact, we achieved very good precision for $\zeta$ ranging the interval $10^{-4} \lapp \zeta \lapp 10^{3}$.

\begin{figure}
\begin{center}
\includegraphics*[width=6cm]{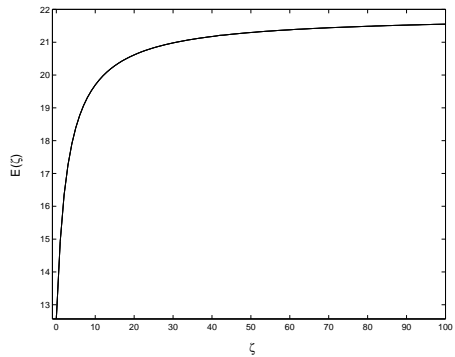}
\end{center}
\caption{
The numerical solution of the energy of the t'Hooft-Polyakov $E(\zeta)$  as a function of 
the ratio of the Higgs to vector field mass, $\zeta$. Note that the energy remains finite when $\zeta \rightarrow \infty$.}
\label{enerkov}
\end{figure}

We also compute the mass of the monopole and verified that it increases monotonically with increasing $\zeta$, remaining finite when $\zeta \rightarrow \infty$ (see Fig. \ref{enerkov}).

\subsection{Bekenstein Model}

Consider a gauge kinetic function, $B_F(\varphi)$, given by: 
\be
B_F(\varphi)\,=\,e^{-\frac{2\varphi}{\sqrt{\omega}}}\, , \label{expo}
\ee
where $\omega$ is a positive coupling constant. This recovers the original 
Bekenstein model\cite{bekenstein}, 

Defining $\psi=\varphi/\sqrt{\omega}$ and using eqn. (\ref{expo}) the 
Lagrangean density in (\ref{laga}) can be written as
\ben
\mathcal{L}\,&=&\, \frac12\,\left(D_\mu\,\Phi^a \right)\left(D^\mu\,\Phi^a\right) - V(\Phi^a) \nn \\
&-& \frac{e^{-2\,\psi}}{4}\,f_{\mu\nu}^{a}\,f^{a\mu\nu}
+\omega\,\partial_\mu\,\psi\,\partial^\mu\,\psi \label{lagabek}.
\een 
Note that despite the fact that the gauge kinetic function in eqn. (\ref{expo}) is only well defined for $\omega > 0$, the model described by the Lagrangean density in eqn. (\ref{lagabek}) allows for both positive and negative values of $\omega$. 


\begin{figure}
\begin{center}
\includegraphics*[width=6cm]{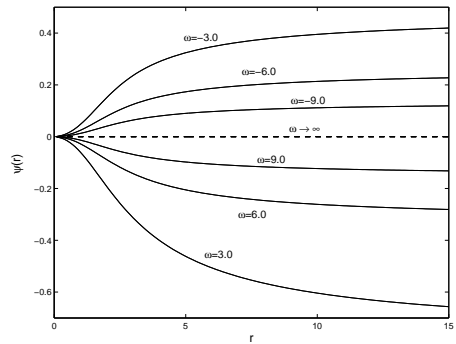}
\end{center}
\caption{The numerical solution of the scalar field 
$\psi \equiv \ln{\epsilon}$ as a function of distance, $r$, 
to the core of the monopole, in the Bekenstein model. 
Note that if $\omega < 0$,  $\epsilon \to \infty$ when $r \to \infty$.
On the other hand, if $\omega > 0$ then $\epsilon \to 0$ when $r \to \infty$.
The dashed line represents the constant-$\alpha$ theory, which corresponds 
to the limit $ \omega \rightarrow \infty$.} 
\label{psibek}
\end{figure}

In this model equations (\ref{V}) and (\ref{VI}) are given by
\ben
\label{Vq}
U'&=&\,2\,\sigma\,U+\frac{W(W^2-1)}{r^2}+e^{2\,\psi}\,H^2\,W\\
\label{VIq}
\sigma'&=&\,-\,\frac{2\,\sigma}{r}-\frac{2}{\omega}\,e^{-2\,\psi}\,\left[\left(\frac{1-W^2}{r^2}\right)^2 + U^2\right]
\een
where we took $\sigma=\psi'$.
\begin{figure}
\begin{center}
\includegraphics*[width=6cm]{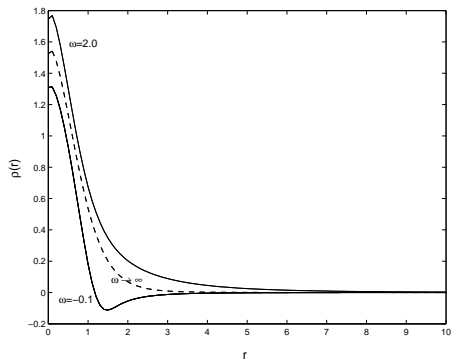}
\end{center}
\caption{The numerical solution of the energy density $\rho$ as a function $r$, 
the distance to the core of the monopole. The dashed-line is for $\omega \rightarrow \infty$. Note that for $\omega=2.0$, the energy density is non negative, while for $\omega=-0.1$ it becomes negative after some distance of the core.} 
\label{rhobek}
\end{figure}
First we note that in the limit $\omega\,\rightarrow\,\infty$ one recovers 
the t'Hooft-Polyakov classical solution described in the previous subsection.  
In  Fig. \ref{psibek} we plot the numerical solution of the scalar field 
$\psi(r) = \ln \epsilon$ for several values of $\omega$. 

From Fig. \ref{psibek} one concludes that if $\omega < 0$ then $\epsilon$ 
diverges asymptotically away from the core of the monopole, and the energy density 
\ben
\rho &=& e^{-2\,\psi}\left[\left(\frac{W'}{r} \right)^2 + \frac12 \left(\frac{1-W^2}{r^2}\right)^2 \right]  \nn \\
&+& \left[ \frac{H'^{2}}{2} + \left(\frac{WH}{r} \right)^2 + \frac18 \zeta^2(1-H^2)^2\, + \omega\,\psi'^{2} \right]
\een
is no longer positive definite. 
However if $\omega > 0$ then $\epsilon$ vanishes asymptotically when $r  \to \infty$, and the energy density is in this case positive definite (See Fig. \ref{rhobek}). We also note by observing Fig. \ref{psibek} that in the large $\omega$ limit the curves for positive and negative 
$\omega$ are approximately symmetric approaching the dashed line which represents 
the constant-$\alpha$ model when $\omega \rightarrow \infty$.

\begin{figure}
\begin{center}
\includegraphics*[width=6cm]{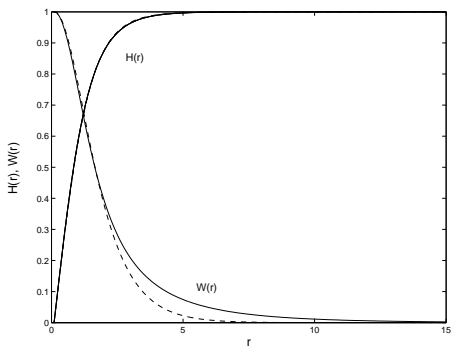}
\end{center}
\caption{The numerical solution of the Higgs field $H(x)$ and the gauge field $W(x)$ as a
function of distance $r$, to the core of the monopole for. Note that the change in $W(r)$ with
respect to the t'Hooft-Polyakov solution(dashed-line) is more relevant than the change in $H(r)$.} 
\label{hwbek}
\end{figure}

Finally, in Fig. \ref{hwbek} we plot the numerical solution of the scalar field $H(r)$ and the gauge field $W(r)$ as a function of distance, $r$, to the core of the monopole. The dashed-line represents the constant-$\alpha$ solution and the solid line represents the Bekenstein one for $\omega=2.0$. 
We verified that even in the $\omega \rightarrow 0$ limit, the change in $H(r)$ with respect to t'Hooft-Polyakov solution is still negligible.


\subsection{Polynomial Gauge Kinetic Function}

We now consider another class of gauge kinetic functions given by
\be
B_F(\varphi)=1.0+\sum_{i=1}^{N}\,\beta_i\,\varphi^i \, , \label{poli}
\ee
where $\beta_i$ are dimensionless coupling constants and $N$ is an integer.

Note that by considering
\be
\beta_i = \frac{(-2)^i}{w^{i/2} i!}\,
\ee 
one recovers the Bekenstein coupling given in (\ref{expo}).
This relationship between the coupling constants $\beta_i$ and $\omega$ has 
interesting consequences for the model given by eqn. (\ref{poli}). 

First we verified that the behaviour of $\psi$ both for $\beta_1 \, >\,0$ and $\beta_1 \, <\,0$  is similar to that of the Bekenstein model with $\omega\,>\,0$ which is recovered in the 
limit of small $|\beta_1|$/large $\omega$. 

Another feature that we verified is that if one takes $\beta_1=0$ one gets the t'Hooft-Polyakov limit, for any of $\beta_i$ for $i>1$. This means that there is a class of gauge kinetic functions for which the
classical static solution is maintained despite the modifications to the model.
 
Another property can be noted when one substitutes the gauge kinetic function in equation of motion for $\varphi$ (\ref{three}). One gets
\ben\label{oddeven}
\frac{1}{r^2}\frac{d}{dr}(r^2\frac{d\varphi}{dr})&=&
C^2\left(\sum_{k=1}^{N}(2k-1)\beta_{2k-1}\varphi^{2k-2}\right)\,
+\nn  \\
&+& C^2 \left(\sum_{k=1}^{N}(2k)\beta_{2k}\varphi^{2k-1}\right)\,
\een
with
\be
C\,=\,\left[\left(W' \right)^2+\frac12\left(\frac{(1-W^2)}{r}\right)^2\right]^{\frac12}. \label{synn}
\ee
Since that $C^2\,>\,0$, when $\beta_i \rightarrow -\beta_i$ one sees by eqn. (\ref{oddeven}) that $\varphi(r)\,\rightarrow\,-\varphi(r)$. However, as $B_F$ in eqn. (\ref{poli}) is kept invariant, $H(r)$ or $W(r)$ do not vary.

Although we have found the set of solutions for several values of $N$, for simplicity we consider that $N=2$ in eqn. (\ref{poli}), that is
\be
B_F = 1.0 + \beta_1 \varphi + \beta_2 \varphi^2, \label{nosso}
\ee
with two free parameters. We define the models $1$ and $2$ as $\beta_1=-3,\,\beta_2=0$ (linear coupling) and 
$\beta_1=-2,\,\beta_2=5$(quadratic coupling) respectively. 

\begin{figure}
\begin{center}
\includegraphics*[width=6cm]{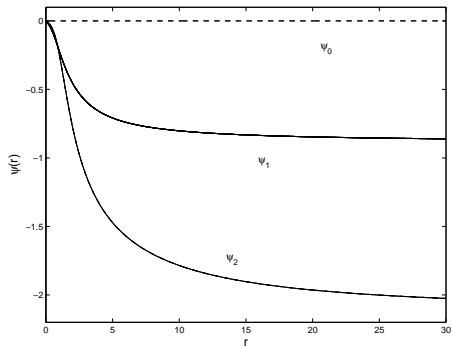}
\end{center}
\caption{The numerical solution of the scalar field 
$\psi \equiv \ln \epsilon$ as a function of distance, $r$, 
to the core of monopole, for a polynomial gauge kinetic function. 
Models 0, 1 and 2 are defined by $\beta_1=0$ ($\beta_2$ arbitrary),  
$\beta_1=-3,\,\beta_2=0$ (linear coupling) and 
$\beta_1=-2,\,\beta_2=5$ respectively.}
\label{psipoli}
\end{figure}
\begin{figure}
\begin{center}
\includegraphics*[width=6cm]{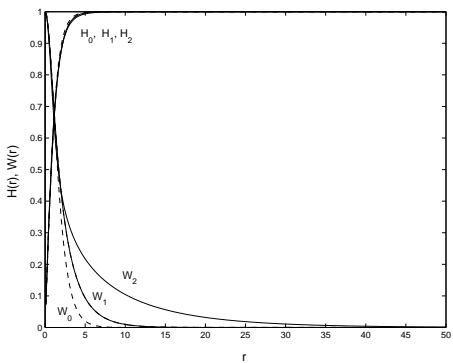}
\end{center}
\caption{
The numerical solution of the fields $H(r)$ and $W(r)$ as a function of 
distance, $r$, to the core of monopole, for models $0$, $1$ and $2$. Note 
that the change in $W(r)$ with respect to the standard 
constant-$\alpha$ result is much more dramatic than the change in $H(r)$.}
\label{hwpoli}
\end{figure}

In Fig. \ref{psipoli} we plot the numerical solution of the scalar field $\psi(r)$ as a function of the radial coordinate. 
As we have shown earlier the model $0$ represents any model with 
$\beta_1 = 0$ and we have verified that the replacement 
$\beta_1 \to -\beta_1$ does not modify the solution for $\psi$.

\begin{figure}
\begin{center}
\includegraphics*[width=6cm]{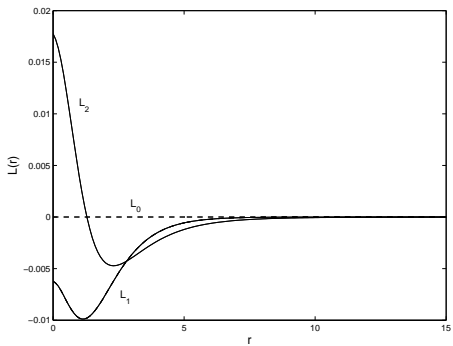}
\end{center}
\caption{Plot of $L_i(r)=\log(H_i/H_0)$ for the models $0$, $1$ and $2$. One clearly sees that even a small value of 
$\beta_1$ leads to a different vortex solution from the standard t'Hooft-Polyakov one.}
\label{lpoli}
\end{figure}

One clearly sees in Fig. \ref{hwpoli} that 
the change in $W(r)$ with respect 
to the standard constant-$\alpha$ result is much more dramatic than the 
change in $H(r)$, we define the function
\be
L_i(r) = \log\left( \frac{H_i}{H_0}\right), \label{zeta}
\ee
and plot in Fig. \ref{lpoli} the results for the different models. 

Note that even a small value of $\beta_1$ will lead to a modification of the magnetic monopole 
solution with respect to the standard t'Hooft-Polyakov solution.

\begin{figure}
\begin{center}
\includegraphics*[width=6cm]{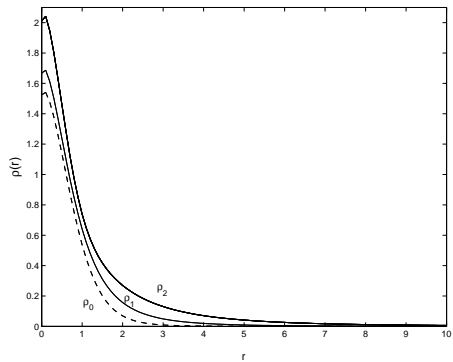}
\end{center}
\caption{The energy density as a function of the 
distance, $r$ to the string core for models 0, 1 and 2. 
The dashed line represents the constant-$\alpha$ model. 
Note that there is an increase of the energy density due to the 
contribution of the extra field $\varphi$ in the case models $1$ and $2$.}
\label{rhopoli}
\end{figure}

We have also studied the behaviour of the energy density in this
model. In Fig. \ref{rhopoli} one sees by comparing with
the dashed line, which represents the constant-$\alpha$ model, 
that since the fine structure constant varies there is a new 
contribution due to the field $\varphi$, to the total energy of the 
topological defect. 
In fact the energy density of the monopole can be divided into two 
components: one that is localized inside the core of the monopole 
and other related to the 
contribution of the kinetic term associated 
with the spatial variations of the fine structure constant.

\section{Constraints on Variations of $\alpha$}

After having discussed in the previous section the numerical solutions for the varying-$\alpha$ monopole, our next investigating point is to find an overall limit of the spatial variations of the fine-structure constant on monopole networks.

For simplicity we assume 
\be
B_F(\varphi)\,=\,1\,+\,\beta\,\varphi, \label{bfl}
\ee 
i.e., the gauge kinetic function is a linear function in $\varphi$ that satisfies the spherically symmetric Poisson equation given by
\be
\nabla^2 \varphi = \frac{1}{4}\, 
\beta \, f^2\,.\label{poisson}
\ee 
Note that $\beta$ is constrained by Equivalence Principle tests to be such that $|\beta| < 10^{-3}\,G^{1/2}$ (see \cite{olive2}).

Integrating eqn.(\ref{bfl}) from  the core up to $r_{max}$, which represents a cosmological cut-off scale, one gets
\be
4 \pi r^2 \frac{d \varphi}{dr}=  \beta I(r)  M(r_{\rm max})\,. \label{poi}
\ee
In eqn.(\ref{poi}) we used the mass of the monopole which is given by
\be
M(r)=4 \pi \int^r_0 \rho(r') r'^2 dr'\,,
\ee
and
\be
I(r)=\frac{\pi}{M(r_{\rm max})} \int_0^r f^2(r') r'^2 dr'\,,
\ee
which is a slowly varying function of $r$ outside the core always smaller than unity. Thus we can take $I(r) \sim {\rm const}$ and integrate eqn. (\ref{poi}) to get
\be
\varphi \sim  \frac{\beta I\,M(r_{\rm max})}{4\pi} \left(\frac1{r}-\frac1{r_0}\right)\,,
\ee
where $r_0$ is a integration constant which could be identified as the core radius. Using $B_F(\varphi)=\epsilon^{-2}$ one gets
\be
\epsilon \sim 1-\frac{\beta^2 I\,M(r_{\rm max})}{8\pi} \left(\frac1{r}-\frac1{r_0}\right)\,,
\label{epsilonr}
\ee
which means that the variation of the fine structure constant away from the
monopole core is proportional to the gravitational potential induced by
the monopoles.

\subsection{GUT Monopoles }

Let us estimate an overall limit for the spatial variation of $\alpha$ outside the core of GUT monopoles. In this context the mass of the monopole is of order of
\be
M(r_{max}) \sim 10^{16}\,GeV.
\ee
The variation of $\alpha$ from the core up to infinity is
\be
\frac{\Delta \alpha}{\alpha} = \frac{\alpha(r \rightarrow \infty)-\alpha(r_0)}{\alpha(r_0)} = \epsilon^2 -1 \lesssim 10^{-13},
\ee
where we have used $\alpha=\alpha_0/B_F(\varphi)$, with $\alpha=\alpha(r \rightarrow \infty)$ and $\alpha_0=\alpha(r_0)$.

\subsection{Planck Monopoles}

Proceeding as above for the Planck scale symmetry with 
\be
M(r_{max}) \sim 10^{19} {\rm Gev}
\ee
one obtains an overall limit for the spatial variation of the fine-structure seeded by magnetic 
monopoles
\be
\frac{\Delta \alpha}{\alpha} \lesssim 10^{-7},
\ee
which is still very small even for Planck scale monopoles.

\section{Conclusion}

In this paper we investigated static monopole solutions in the context of varying-$\alpha$ theories 
based on Bekenstein-type models.
First we studied various models with constant $\alpha$ and reviewed the standard static t'Hooft-Polyakov magnetic monopole 
solution. 
Then we considered models with varying-$\alpha$ and showed that despite of the existence of a class of models for which the t'Hooft-Polyakov standard solution is still valid, in general, our solutions depart from the former one. We showed that Equivalence Principle 
constraints impose tight limits on the variations of $\alpha$ induced by
magnetic monopoles. This confirms 
the difficulty to generate significant large-scale spatial variation of 
the fine structure constant found in previous works, even in the most favorable case 
where these variations are seeded by magnetic monopoles.

\section*{acknowledges}
J. Menezes was supported by a Brazilian Government grant - CAPES-BRAZIL 
(BEX-1970/02-0).  
Additional support came from Funda{\c c}\~ao para a Ci\^encia e a Tecnologia 
(Portugal) under contract POCTI/FP/FNU/50161/2003.

\end{document}